\documentclass[journal]{IEEEtran}
\usepackage{cite}
\usepackage{amsmath,amssymb,amsfonts}
\usepackage{algorithmic}
\usepackage{graphicx}
\usepackage{tabularx}
\usepackage{adjustbox}
\usepackage{svg}
\usepackage{multirow}
\usepackage{algorithm}
 \usepackage{booktabs}
\usepackage{soul}
\usepackage{xparse}
\usepackage{textcomp}
\usepackage{stfloats}
\usepackage[caption=false,font=normalsize,labelfont=sf,textfont=sf]{subfig}
\begin{document}
\title{A Comparative Study of EMG- and IMU-based Gesture Recognition at the Wrist and Forearm}
\author{Soroush Baghernezhad, \IEEEmembership{Student Member, IEEE}, Elaheh Mohammadreza, \IEEEmembership{Student Member, IEEE}, Vinicius Prado da Fonseca, \IEEEmembership{Member, IEEE}, Ting Zou, \IEEEmembership{Senior Member, IEEE}, and Xianta Jiang, \IEEEmembership{Senior Member, IEEE} 

\thanks{\textcolor{red}{
This work has been submitted to the Springer Nature Link for possible publication.\\
Copyright may be transferred without notice, after which this version may
no longer be accessible.}\\
Submitted on October 28th, 2025. This work was supported in part by the Government of  Canada's New Frontiers in Research Fund (NFRF, Grant No NFRFE-2022-00407), Natural Sciences and Engineering Research Council of Canada's Discovery Grant (DGECR-2020-00296), and Natural Sciences and Engineering Research Council of Canada's Research Tools and Instruments (NSERC RTI, Grant No  RTI-2022-00688).}
\thanks{This work involved human subjects or animals in its research. Approval of all ethical and experimental procedures and protocols was granted by the Memorial University Interdisciplinary Committee on Ethics in Human Research (20210316-SC).}
\thanks{
Soroush Baghernezhad, Elaheh Mohammadreza, Vinicius Prado da Fonseca, and Xianta Jiang are with the Department of Computer Science, Memorial University of Newfoundland, St. John’s, NL A1B 3X5, Canada. (email: sbaghernezha@mun.ca, emohammadrez@mun.ca, vpradodafons@mun.ca, xiantaj@mun.ca).
}
\thanks{Ting Zou is with the Department of Mechanical and Mechatronics Engineering, Memorial University of Newfoundland, St. John’s, NL A1B 3X5, Canada. (email: tzou@mun.ca). }
}
\markboth{}
{Baghernezhad \MakeLowercase{\textit{et al.}}: A Comparative Study of EMG and IMU-based HGR}

\maketitle
\begin{abstract}
Gestures are an integral part of our daily interactions with the environment. Hand gesture recognition (HGR) is the process of interpreting human intent through various input modalities, such as visual data (images and videos) and bio-signals. Bio-signals are widely used in HGR due to their ability to be captured non-invasively via sensors placed on the arm. Among these, surface electromyography (sEMG), which measures the electrical activity of muscles, is the most extensively studied modality. However, less-explored alternatives such as inertial measurement units (IMUs) can provide complementary information on subtle muscle movements, which makes them valuable for gesture recognition. In this study, we investigate the potential of using IMU signals from different muscle groups to capture user intent. Moreover, we compare different muscle groups and check the quality of pattern recognition on individual muscle groups. Our results demonstrate that IMU signals capture unique information not reflected in sEMG signals and can serve as an effective input modality for static gesture recognition. IMU is demonstrated as a promising modality for controlled static gesture tasks and a useful alternative when EMG acquisition is inconvenient. This approach also offers new possibilities for hand gesture recognition in fields such as robotics, prosthetics, teleoperation, sign language interpretation, and beyond.
\end{abstract}


\section{Introduction}
\label{sec:introduction}

\IEEEPARstart{H}{and} gestures are intentional motions of the hands and fingers that play an important role in human interactions \cite{SONG2021298}, from expressing an emotion to conveying a meaning. Hand gestures are also a great means of human and computer interaction (HCI) and have been used in a wide range of applications, including text input systems \cite{7926343}, air gesture control \cite{jiang_feasibility_2018}, prosthetic control \cite{jiang_emerging_2022}, sign language recognition \cite{shin_methodological_2024}, rehabilitation\cite{connolly_imu_2017, liu2016development}  etc.
\begin{figure}[!ht]
\centerline{\includegraphics[width=\columnwidth]{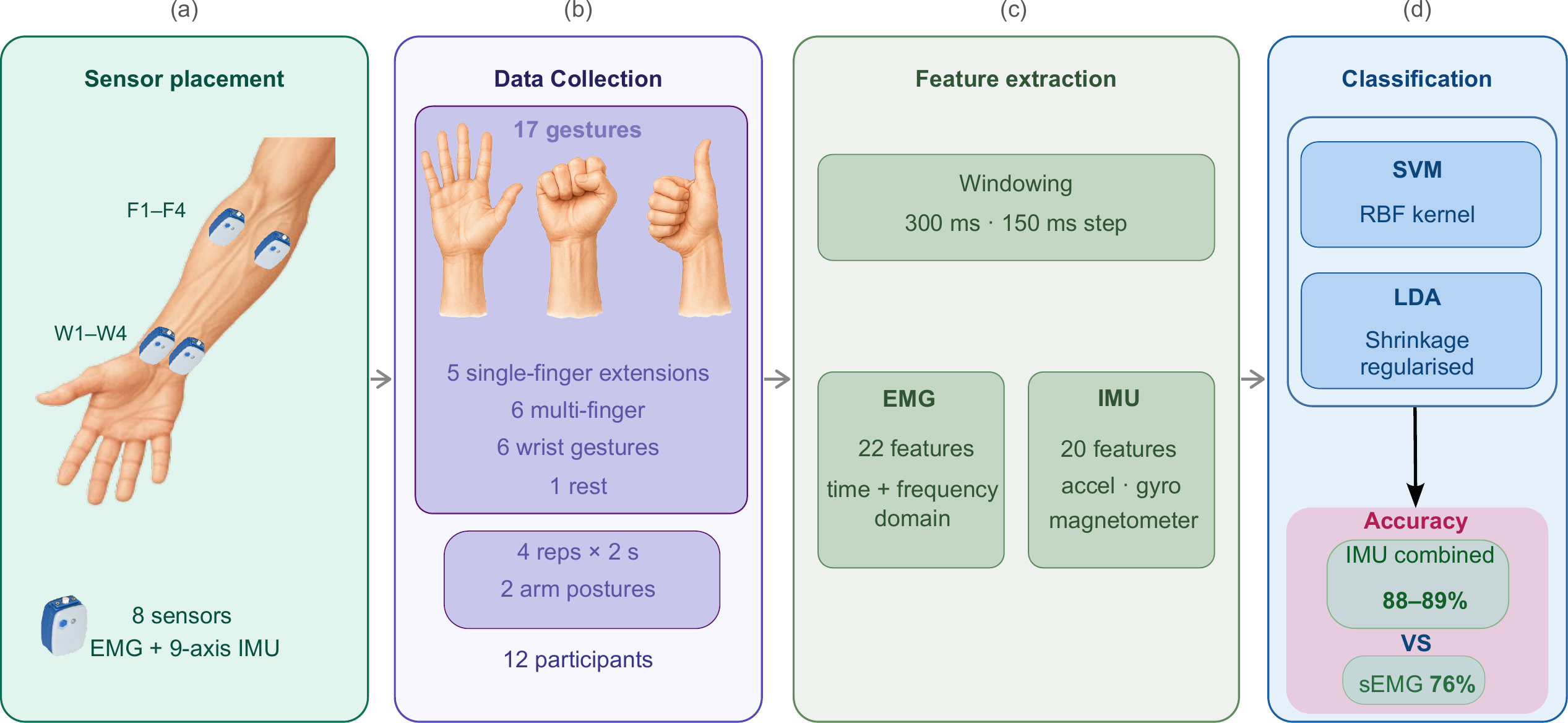}}
        
    \caption{Overview of our work. (a) Sensor Placement (b) Gestures (c) Feature Extraction (d) Model Training}

    \label{fig:one}
\end{figure}
The formation of hand gestures begins in the brain, where signals are generated to control movements. These signals travel through the nervous system to the muscles in the arm and wrist, activating them to contract. This contraction results in tendon movements, skin and blood vessel deformation, and ultimately, hand gestures. These changes can be captured and used as input for pattern recognition systems \cite{shin_methodological_2024}. \\

Researchers have used wearable-sensor and vision-based approaches to acquire inputs for gesture recognition algorithms. Vision-based gesture recognition (VGR) approaches face several challenges, such as complex backgrounds that hinder hand segmentation, variations in illumination, hand movement, and occlusion of the hand gestures \cite{https://doi.org/10.1049/iet-cvi.2017.0052}. On the other hand, sensor-based approaches have been studied extensively, particularly electrical sensors such as sEMG, which is the most commonly used modality in the literature\cite{shin_methodological_2024}. However, sEMG signals are ineffective for detecting tendon movements \cite{10136753}. Furthermore, the performance of sEMG-based systems is highly dependent on the placement of sensors on the arm or forearm, yet only a few studies have focused on optimizing sensor placement \cite{botros_electromyography-based_2022,electronics13153072,HOGLUND20211}. Inertial sensors, on the other hand, have been mostly used for human activity and gait detection, and not for static gesture recognition. A few studies have used IMU gloves for HGR\cite{informatics5020028,10069798}; however, these gloves have several drawbacks, including the need for calibration and adjustment for different hand sizes and more importantly, they are not usable for amputees. Although IMUs have been used in combination with EMG \cite{s22041321,jiang_feasibility_2018} for HGR, the potential of using inertial sensors alone, as well as their optimal placement, remains unexplored.

The goal of this study is to compare the performance of EMG and IMU sensors across different placements on the arm. To achieve this, eight EMG and IMU sensors were placed on different muscle groups to identify the optimal sensor locations. Each sensor recorded surface electromyography, accelerometer, gyroscope, and magnetometer data. The signals were then preprocessed and classified using various machine learning classifiers. Figure \ref{fig:one} provides an overview of our work.

Our main contributions are as follows:
\begin{itemize}
\item Demonstrated the potential of IMU signals for static gesture classification.
\item Identified the optimal placement of IMU sensors on the arm to enhance classification performance.
\item Verified the performance of IMU in capturing static grasps across four different sensor placements and two postures, compared to sEMG.
\end{itemize}

The remainder of this paper is organized as follows. Section \ref{sec:litrev} reviews existing studies on hand gesture recognition, highlighting the advantages and limitations of various approaches. Section \ref{sec:method} details our proposed methodology, including data collection, preprocessing and signal evaluation, feature extraction, and classification. Section \ref{sec:results} presents the experimental results, followed by a discussion where we analyze the findings and their implications. Section \ref{sec:conclusion} concludes the paper and highlights the main findings. Finally, Section \ref{sec:limitations} outlines potential directions for future research.

\section{Literature Review}
\label{sec:litrev}


In wearable-sensor approaches, one or more sensors are placed on the wrist \cite{jiang_feasibility_2018,botros_electromyography-based_2022}, arm \cite{botros_electromyography-based_2022}, back of the hand, or fingers \cite{connolly_imu_2017}. These sensors can be electrical, such as EMG \cite{JIANG201763} and electrical impedance tomography (EIT); mechanical, such as forcemyography (FMG) \cite{7908995,continuousprediction} and inertial measurement units (IMU); optical, such as photoplethysmography (PPG); or acoustical, such as mechanomyography (MMG) \cite{jiang_emerging_2022}.  

In previous studies, researchers have utilized different input modalities for pattern recognition, with sEMG being the most commonly used sensor in the literature. The authors in \cite{botros_electromyography-based_2022} demonstrated that wrist sEMG signals are effective for both single- and multi-gesture recognition, achieving an average accuracy of 84.5\% on 17 gestures with a selected set of features. In this study, they exclusively used sEMG sensors and collected data from a fixed arm position. The authors in \cite{lee2021} used three sEMG sensors on the forearm and achieved an accuracy of 94\% using an artificial neural network (ANN) on 10 gestures. However, their accuracy decreased when the number of gestures increased. The authors in \cite{s20143994} used an RNN with GRU cells to predict 20 hand gestures in real time, achieving an accuracy of 89.6\% using a 200 ms window of data.

IMUs have also been used in the literature for HGR, Motion and gait analysis \cite{s20154345}. The authors in \cite{connolly_imu_2017} used 16 IMUs on a data glove to measure the range of motion in rheumatoid arthritis patients. In \cite{informatics5020028}, the authors implemented five IMUs on the fingertips of a data glove. They used a random forest (RF) classifier to classify 26 gestures in French Sign Language, achieving an overall accuracy of 92.4\%. The authors in \cite{10069798} also used a data glove with three IMU sensors mounted on the index, thumb, and middle fingers and employed an LSTM combined with an ANN to classify 11 gestures. Kefer et al. \cite{10.5555/3177188.3177192} evaluated whether IMU data from the wrist could outperform data from the forearm for dynamic gesture classification. However, they found no significant difference between the classification accuracies of IMUs placed on the wrist and forearm.

Sensor fusion is another technique that has been used to improve the accuracy of classifiers. The authors in \cite{s22041321} used sensor fusion techniques to leverage both EMG and IMU data for training user-independent models to classify seven different gestures. They achieved a mean accuracy of 84.6\% when using both EMG and IMU. Jiang et al. \cite{jiang_feasibility_2018} fused both IMU and sEMG data to classify eight air and surface gestures using linear discriminant analysis (LDA). They achieved an accuracy of 92.8\% and suggested that wrist-worn bands are helpful for real-time gesture recognition. However, the feasibility of IMU-only wristband devices was not studied in their work. Krasoulis et al. \cite{Krasoulis2017} integrated sEMG with IMU data to decode 40 distinct movements and showed that the inclusion of IMU data significantly improved classification accuracy. However, their study did not investigate sensor placement at the wrist.

Some studies investigate the optimal placement of sensors. In \cite{Jarque-Bou2018}, the authors addressed the challenge of finding a stable sensor placement on the hand while preserving informative signals. They reduced the number of sEMG electrodes from 30 to only 7. The authors in \cite{electronics13153072} also used sEMG to evaluate optimal placement of sensors for prosthetic control and suggested that placing two sensors on the wrist yields the best accuracy. Authors in \cite{HOGLUND20211} investigated the placement of IMU sensors for assessing upper limb motion and suggested the optimal positions for placing inertial sensors on the forearm, upper arm, and scapula. Studies related to gesture classification are summarized in table \ref{tab:comparison}.
\begin{table*}[!ht]
\centering
\caption{Comparison of sensor-based HGR studies}
\label{tab:comparison}
\renewcommand{\arraystretch}{1.4} 
\begin{adjustbox}{width=\textwidth}

\begin{tabular}{p{2.8cm} p{1.2cm} p{1.8cm} p{2.2cm} p{1.4cm} p{1.2cm} p{1.8cm} p{1.5cm}}
\hline
\textbf{Work} & \textbf{Sensor} & \textbf{Placement} & \textbf{Muscles} & \textbf{Gesture Type} & \textbf{No. of Gestures} & \textbf{Method} & \textbf{Accuracy} \\
\hline
Botros et al.\cite{botros_electromyography-based_2022} & sEMG & wrist and forearm & FCU, ED, ECU, EDM, BRR, FCR, PL, FDS & static & 17 & LDA-SVM & 84.5 \\
Lee et al.\cite{lee2021} & sEMG &  forearm & FCU, FCR, BR & static & 10 & ANN & 94 \\
Zhang et al.\cite{s20143994} & sEMG &  forearm & - & static & 20 & RNN & 89.6 \\
Kim et al. \cite{electronics12071541} & sEMG &  forearm & - & dynamic & 10 & CRNN & 96.04 \\
Mummadi et al. \cite{informatics5020028} & IMU &  hand & - & static & 22 & RF & 92.4 \\
Lauss et al. \cite{10069798} & IMU &  hand & - & - & 11 & LSTM + ANN & 93 \\
Kefer et al. \cite{10.5555/3177188.3177192} & IMU &  wrist and forearm & - & dynamic & 8 & KNN, RF, DT &  83.52 to 94.44 \\
Alfaro et al. \cite{s22041321} & sEMG, IMU &  forearm & ECU & static & 7 & Adaptive LS-SVM & 92.4 \\
Jiang et al. \cite{jiang_feasibility_2018} & sEMG, IMU &  wrist & FDS, BR, FCU, ED & air and surface & 8 & LDA & 92.8 \\
\hline
\end{tabular}
\end{adjustbox}
\end{table*}

\noindent

\section{Methodology}
\label{sec:method}

\subsection{Sensor Placement}
For this study, Noraxon Ultium system is used to capture sEMG and IMU data. Each sensor unit contains both sEMG and IMU sensors, and 8 units were used in this study. sEMG signals are recorded at 2000 Hz, and IMU is recorded at 200 Hz.
Noraxon MR3 software was employed for data collection.

\begin{figure}[!ht]
    \centering
    \centerline{\includegraphics[width=\columnwidth]{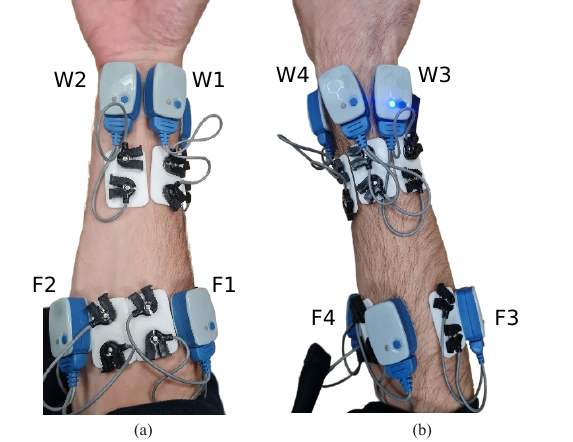}}
    
    \caption{Placement of sensors on (a) Anterior and (b) Posterior side of the arm.}
    \label{fig:placement}

\end{figure}

As suggested by \cite{konrad2006abc}, we placed our sensors parallel to the muscle fibers. The muscle groups were selected to ensure that the primary muscles responsible for the desired hand gestures were targeted \cite{frank2019atlas}.  
Four sensor units (sEMG + IMU) are being placed circumferentially around the wrist (W1-W4), and the other four were placed circumferentially around the forearm (F1-F4) as shown in Figure \ref{fig:placement} . Our sensor placements are as follows (for the left arm):

\begin{itemize} 
\item F1: Positioned on the brachioradialis (BR), with the potential to capture motions of the deeper layers, including the pronator teres (PT) and extensor carpi radialis longus (ECRL).
\item F2: Placed on the palmaris longus (PL) and flexor carpi radialis (FCR).
\item F3: Positioned on the right side of the flexor digitorum profundus (FDP), flexor carpi ulnaris (FCU), and flexor digitorum superficialis (FDS) (opposite of F2), capturing signals from the deep layers.
\item F4: Placed on the left side of the ulna, over the extensor digitorum (ED) and extensor carpi radialis brevis (ECRB) (opposite of F1), with remote capture of the ECRL and extensor carpi ulnaris (ECU).
\item W1: Positioned over the flexor pollicis longus (FPL), part of the pronator quadratus (PQ), and the FCR tendon.
\item W2: Placed over the PL tendon, flexor digitorum superficialis (FDS) muscle, and tendons, with the ability to also capture signals from the FDP.
\item W3: On ECU and ED tendons.
\item W4: On EPL, ECRB, and EPB tendons.
\end{itemize}
Before sensor placement, the participants' arms were cleaned with rubbing alcohol to improve skin impedance \cite{konrad2006abc}. The forearm and wrist were then examined while performing simple gestures to identify the corresponding muscles. Once the target muscles were located, the muscle group locations were marked using a cosmetic-grade, skin-safe marker. Subsequently, sensors were carefully placed on the marked locations.

\subsection{Data Collection}
For this study, data were collected from 12 able-bodied participants (age: $26 \pm 3$ years, 4 females, 8 males) with an average wrist circumference of 17.5 $\pm$ 2~cm, forearm circumference of $25.8 \pm 3 $~cm, and forearm length of $25 \pm 1.5$~cm. All participants had no prior neuromuscular disorders. Prior to the experiment, they were given a consent form approved by Memorial University of Newfoundland Interdisciplinary Committee on Ethics in Human Research.

\begin{figure}[!ht]
    \centering
     \includegraphics[scale=0.8]{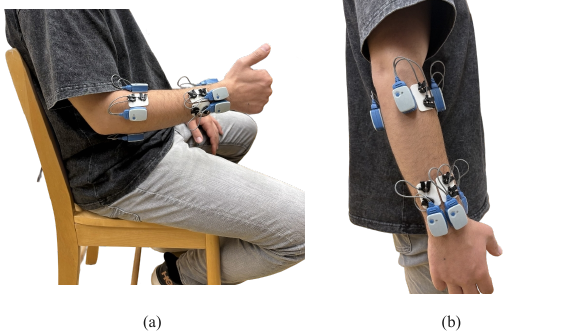}   
\caption{(a) Participant postures during phase one  and (b) phase two of the experiment.}

    \label{fig:twopos}
\end{figure}

After securely placing the sensors on the participants’ forearm and wrist, data collection was carried out in two phases. In phase one, participants were seated comfortably with their elbows bent at a 90-degree angle and their hands resting, as illustrated in Figure \ref{fig:twopos}. In phase two, participants stood naturally with their arms positioned alongside their bodies. In both phases, visual instructions were displayed on a screen positioned in front of the participants. Before starting phase one, participants were asked to hold their hand in a stationary resting position for 15 seconds to collect the calibration data to be used in the signal analysis.

\begin{figure}[!ht]
    \centering
     \includegraphics[scale=.5]{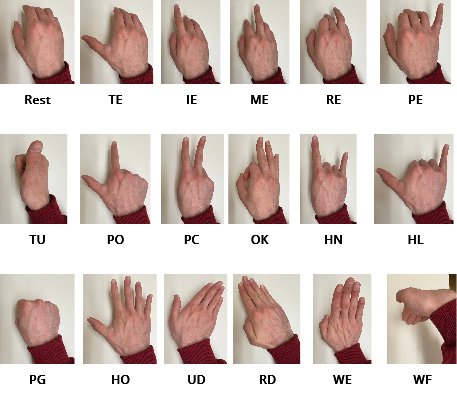}   
\caption{Gestures used in this experiment. From top left: Rest, Thumb Extension (TE), Index Extension (IE), Middle Extension (ME), Ring Extension (RE), Pinky Extension (PE), Thumbs Up (TU), Pointing (PO), Peace (PC), OK, Horn (HN), Hang Loose (HL), Power Grasp (PG), Hand Open (HO), Ulnar Deviation (UD), Radial Deviation (RD), Wrist Extension (WE), and Wrist Flexion (WF).}

    \label{fig:gestures}
\end{figure}

Prior to recording, participants were asked to practice the entire set of gestures once. During the experiment, they followed the visual instructions and performed 17 distinct gestures which included contractions for five single-finger extensions, six multi-finger movements, and six conventional wrist gestures that are practical in industrial applications and daily tasks \cite{botros_electromyography-based_2022}. as shown in Figure \ref{fig:gestures}. Each gesture was repeated four times, with each repetition lasting 2 seconds, followed by a 2-second rest period. Additionally, a 5-second rest was provided before the start of each new gesture to avoid muscle fatigue.
After completing all 17 gestures, participants were allowed to rest before proceeding to the next position.

The raw data were labeled using an automatic labeling method based on visual cues provided to the participants. Previous studies have shown that there is no significant difference in classification accuracy between automatic labeling and manual labeling based on video recordings of the experiment \cite{10942474}. EMG activation peaks were aligned with the corresponding labels.

\subsection{Preprocessing}
We collect data at 2000 Hz for EMG and 200 Hz for IMU, which is then upsampled to 2000 Hz.  
Each gesture is repeated 4 times, yielding 16 seconds of data per gesture.  
Our dataset consists of a total of 10,880,000 samples per participant.

\begin{figure*}[!ht] 
    \centering
    \includegraphics[scale=1]{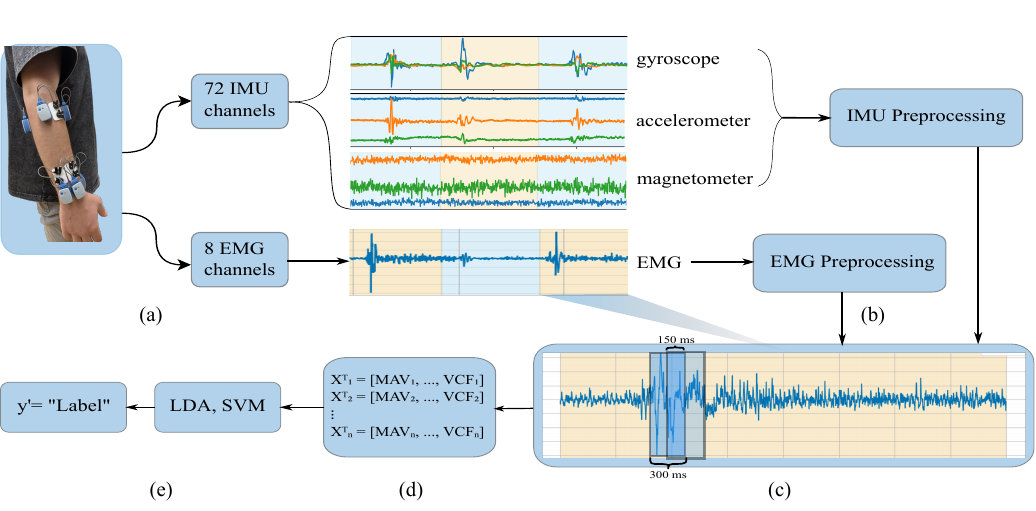} 
    \caption{Overview of the general pipeline used in this study. (a) Data collection (b) Filtering and preprocessing (c) Windowing (d) Feature extraction (e) Classification and Prediction}
    \label{fig:pipeline}
\end{figure*}

\begin{equation}
    \begin{aligned}
        n_{\text{samples}} &= (\text{duration}_{\text{gesture}} + \text{duration}_{\text{rest}}) \cdot n_{\text{repetitions}} \\
        &\cdot n_{\text{gestures}} \cdot n_{\text{channels}} \cdot \text{frequency}
    \end{aligned}
\end{equation}
The raw EMG signals were first filtered using a notch filter to remove powerline noise. Subsequently, a band-pass filter between 20 Hz and 500 Hz was applied. The data were then smoothed using a window length of 50 and detrended to eliminate baseline drift. Smoothing was also applied to the IMU signals.

A sliding window of 300 ms with a 150 ms overlap is used to segment the data, providing sufficient data while maintaining real-time performance \cite{204774}. After segmentation, a set of 22 time-domain and frequency-domain features is extracted from the EMG signals, while the same set of features, excluding MYOP and WAMP, is extracted from the IMU signals. These excluded features are threshold-based, and their thresholds were defined in previous EMG studies; therefore, they are not meaningful for IMU signals. Table \ref{tab:extracted_features} shows the list of extracted features. These features are among the most commonly used in the literature \cite{PHINYOMARK20134832, jiang_feasibility_2018, botros_electromyography-based_2022}.

\begin{table}[htbp]
\centering
\caption{List of Extracted Features \cite{botros_electromyography-based_2022} }
\label{tab:extracted_features}
\begin{tabular}{lll}
\hline
\textbf{Type} & \textbf{Feature Name} & \textbf{Abbreviation} \\
\hline
\multirow{12}{*}{Time Domain} 
& Mean Absolute Value & MAV \\
& Variance  & VAR \\
& Root Mean Square & RMS \\
& Waveform Length & WL \\
& Difference Absolute Mean Value & DAMV \\
& Difference Absolute STD & DASDV \\
& Zero Crossing & ZC \\
& Myopulse Percentage Rate & MYOP \\
& Willison Amplitude & WAMP \\
& Slope Sign Change & SSC \\
& Histogram, 10-bins & HIST \\
& AR Coefficients, 4\textsuperscript{th} Order & AR \\
\hline
\multirow{10}{*}{Frequency domain} 
& Mean Frequency & MNF \\
& Median Frequency & MDF \\
& Peak Frequency & PKF \\
& Total Power & TTP \\
& 1\textsuperscript{st} Spectral Moment & SM1 \\
& 2\textsuperscript{nd} Spectral Moment & SM2 \\
& 3\textsuperscript{rd} Spectral Moment & SM3 \\
& Frequency Ratio & FR \\
& Power Spectrum Ratio & PSR \\
& Variance of Central Frequency & VCF \\
\hline
\end{tabular}
\end{table}

\noindent

Since IMU signals are highly non-stationary and fluctuate rapidly, we require features that can quantify the overall signal energy and capture these fluctuations. 

In this study, we use a modified set of time-domain features proposed in \cite{botros_electromyography-based_2022}, which are designed to capture the signal attributes. We exclude features that are exclusively related to EMG by identifying those that rely on threshold-based computations.

\subsection{Classification}
To classify the gestures, we feed the features from the previous step into both Linear Discriminant Analysis (LDA) and Support Vector Machine (SVM) classifiers. We chose these two conventional machine learning models because they have been shown to be effective for gesture classification \cite{s22041321, jiang_feasibility_2018}, especially when compared to more advanced sequential models such as CRNN \cite{electronics12071541}. 

We use grid search for hyperparameter tuning, where \texttt{shrinkage} controls the regularization strength and \texttt{tol} sets the convergence tolerance in LDA, while in SVM, \texttt{C} regulates the trade-off between margin size and classification error, and the kernel type defines the mapping of data into higher-dimensional space.
 Model performance is evaluated using accuracy as the primary metric. We adopt a 4-fold stratified cross-validation strategy, where one repetition is held out as the test set in each fold. The final accuracy is computed by averaging the results across all folds. Figure \ref{fig:pipeline} shows the overview of our pipeline.

\subsection{Statistical analysis}
This study aims to evaluate whether EMG signals provide a better performance compared to inertial sensing modalities over different sensor placements. We therefore defined four hypotheses: the first three address comparisons between EMG and each individual IMU component (accelerometer, gyroscope, and magnetometer), while the fourth assesses EMG relative to the combined contribution of all IMU components. These hypotheses are as follows:

\begin{align}
\text{H}^{(1)} &: \mu_{\text{EMG}} \geq \mu_{\text{accel}} \\
\text{H}^{(2)} &: \mu_{\text{EMG}} \geq \mu_{\text{gyro}} \\
\text{H}^{(3)} &: \mu_{\text{EMG}} \geq \mu_{\text{mag}} \\
\text{H}^{(4)} &: \mu_{\text{EMG}} \geq \mu_{\text{IMU\_combined}}
\end{align}

To assess the statistical significance and effect size between the two groups, we employed both hypothesis testing and effect size calculation. Normality of each group was first tested using the Lilliefors test \cite{Lilliefors}. If both distributions were found to be normal (\( p > 0.05 \)), we used an independent two-sample t-test to compute the p-value. If at least one of the distributions deviated from normality, the Wilcoxon signed-rank test was applied instead. Additionally, Cohen’s \( d \) was computed to quantify the effect size between the two groups, providing an interpretable measure of the magnitude of difference regardless of statistical significance.\\

\section{Results}
\label{sec:results}

\subsection{Classification results}
 
\begin{figure*}[!ht] 
    \centering
    \includegraphics[scale=1]{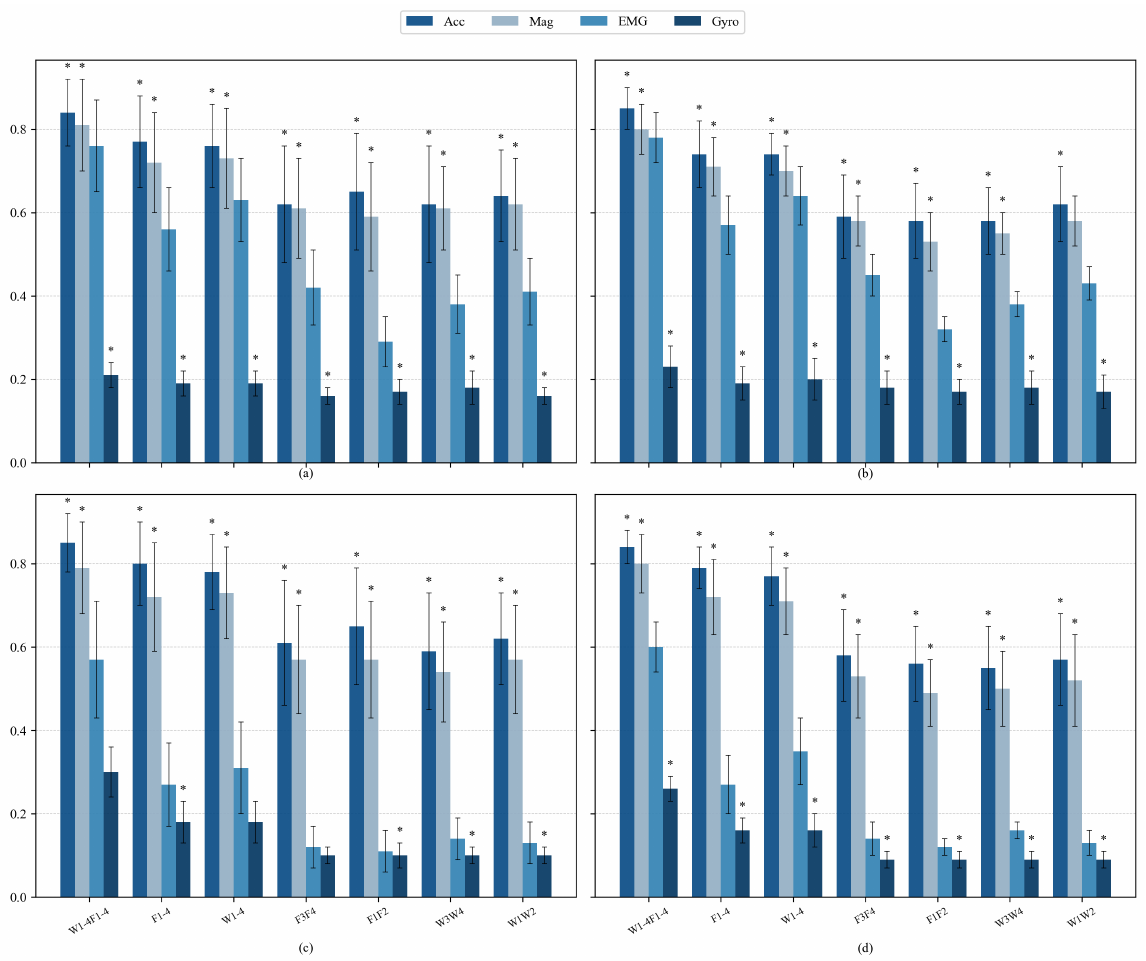}    
    \caption{ Statistical Results for Classification accuracies EMG vs IMU sensors. (a) Comparison of EMG and IMU Sensors Using LDA Classifier at 90° Position, (b) LDA Classifier at 180° Position, (c) SVM Classifier at 90° Position, (d) SVM Classifier at 180° Position. Statistical significance is indicated by * ($p<0.05$ )}
    \label{fig:tables}
\end{figure*}

\begin{figure*}[htbp] 
    \centering
    \includegraphics[scale=1]{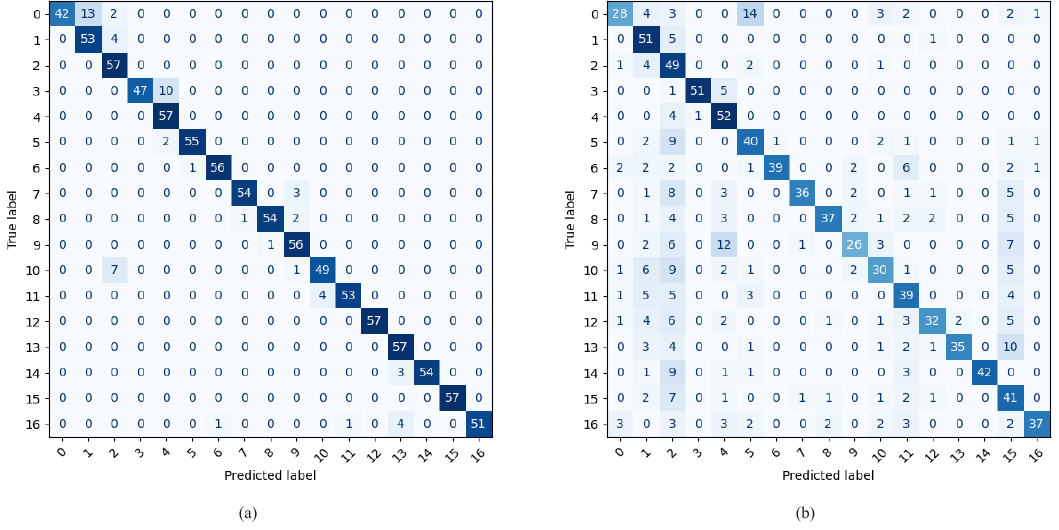}    
    \caption{Confusion matrix for all gestures of a single fold, from one participant. (a) IMU and (b) EMG . 0:TE 1:IE 2:ME 3:RE 4:PE 5:TU 6:RA 7:PO 8:OK 9:H 10:HL 11:PG 12:HO 13:WE 14:WF 15:UD 16:RD }
    \label{fig:confusion}
\end{figure*}

In Figure \ref{fig:tables}, the average accuracies of different sensors across various placements and postures are illustrated. It can be observed that using only acceleration sensor features consistently outperforms EMG features across all placements. SVM did not achieve satisfactory performance with EMG features overall. It is noteworthy that the random accuracy for 17 classes is approximately 5.9$\%$. Both EMG and accelerometer sensor results exhibit high standard deviations, with EMG showing an average accuracy of $76 \pm 11\%$ and the accelerometer showing $84 \pm 8\%$. likely due to variations in participants’ skin conditions, muscle mass, and contraction effort. Based on these results, the first hypothesis is rejected, as EMG did not outperform the accelerometer.

Changing the position of the arm from 90 degrees to 180 degrees does not substantially affect the trend in results, as shown in Figure \ref{fig:tables} (b): in both 90-degree and 180-degree positions, acceleration sensor features consistently outperform EMG features across all placements. This suggests that acceleration-based features are robust to changes in arm position, maintaining superior classification accuracy regardless of posture.

    Gyroscope sensor, however, could not classify the gestures with high accuracy. As shown in Figure~\ref{fig:tables}(a) and Figure~\ref{fig:tables}(b), the EMG sensor consistently outperforms the gyroscope across all placements, which supports the second hypothesis. This can be attributed to the gyroscope’s sensitivity to rotational velocity, which is largely absent in static gestures.

We can also see that in all subfigures of Figure \ref{fig:tables}, magnetometer-based features outperform EMG features across all sensor placements and remain robust to changes in posture. Furthermore, the classification accuracies achieved by magnetometers are comparable to those of accelerometers, assuming no magnetic interference in the environment. The magnetometer captures the ambient magnetic field along multiple axes, and as the sensor moves, the recorded signals vary according to the direction of movement. These signal variations, even in response to subtle muscle movements, form meaningful patterns that can be effectively leveraged for static gesture classification. Therefore, our third hypothesis is rejected.

\begin{figure}[!ht]
    \centering
     \includegraphics[scale=1]{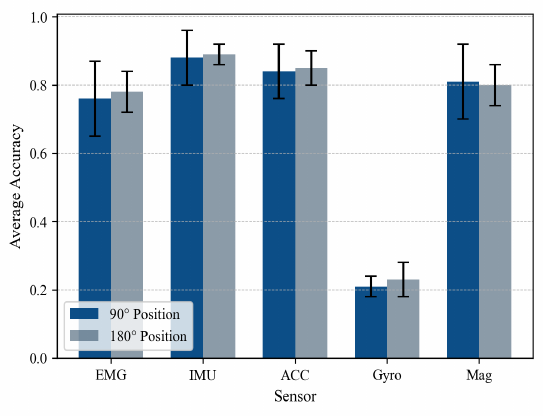}
\caption{Average accuracy of different sensor types on all placements.}

    \label{fig:accuracy_comparison}
\end{figure}

Combining all features from the IMU sensors yields the highest classification performance, achieving an accuracy of $88.0 \pm 8.0\%$ when using data from all sensor placements in the $90^\circ$ arm posture, and $89.0 \pm 3.0\%$ in the $180^\circ$ arm posture. Detailed results for individual sensor placements are presented in Tables \ref{tab:imu90} and \ref{tab:imuleg}. The high effect sizes observed in both classifiers indicate a significant difference between the two groups. Additionally, we calculated the Davies–Bouldin Index (DBI), which measures clustering quality by evaluating the ratio of intra-cluster (within-cluster) dispersion to inter-cluster (between-cluster) separation; lower DBI values indicate better clustering performance. Based on these results, we reject our fourth hypothesis as combined IMU components consistently outperform EMG. Figure \ref{fig:accuracy_comparison} summarizes the classification results.

The confusion matrices of two models: LDA for IMU (left) and LDA for EMG (right) are shown in Figure~\ref{fig:confusion}. The performance of the EMG model decreases as the gestures become more complex, whereas the IMU model maintains consistent performance across different gestures. Moreover, the EMG model often confuses the TE gesture with TU, likely because both involve thumb extension, while TU additionally includes wrist rotation. In contrast, the IMU-based model appears more robust to such confusions.

\begin{table*}[htbp]
\centering
\caption{Statistical Results for Classification accuracies EMG vs (Accelerometer + Gyroscope + Magnetometer) in 90 degree position.}
\begin{adjustbox}{width=\textwidth}
\begin{tabular}{ l c c c c c c c c c c c c }
\toprule
 \textbf{placement} & \multicolumn{4}{c}{\textbf{LDA}} & \multicolumn{4}{c}{\textbf{SVM}} & \multicolumn{4}{c}{\textbf{DBI}} \\
\cmidrule(lr){2-5} \cmidrule(lr){6-9} \cmidrule(lr){10-13} 
&EMG & IMU & $p$ & $d$ & EMG & IMU & $p$ & $d$ & EMG & IMU & $p$ & $d$  \\
\midrule
W1W2 & 0.41(0.08) & 0.74(0.10) & * & 3.68 & 0.13(0.05) & 0.74(0.09) & * & 8.08 & 6.28(1.29) & 2.43(0.98) & * & 3.36 \\
W3W4 & 0.38(0.07) & 0.73(0.11) & * & 3.66 & 0.14(0.05) & 0.73(0.10) & * & 7.33 & 4.66(1.84) & 2.19(0.94) & * & 1.69 \\
F1F2 & 0.29(0.06) & 0.76(0.11) & * & 5.20 & 0.11(0.05) & 0.76(0.10) & * & 8.10 & 5.13(2.18) & 2.12(1.03) & * & 1.77 \\
F3F4 & 0.42(0.09) & 0.73(0.10) & * & 3.24 & 0.12(0.05) & 0.76(0.10) & * & 7.82 & 5.30(1.71) & 2.26(0.80) & * & 2.28 \\
W1-4 & 0.63(0.10) & 0.83(0.09) & * & 2.04 & 0.31(0.11) & 0.82(0.08) & * & 5.28 & 3.88(0.62) & 2.23(0.97) & * & 2.03 \\
F1-4 & 0.56(0.10) & 0.83(0.09) & * & 2.83 & 0.25(0.10) & 0.83(0.09) & * & 6.21 & 4.40(1.24) & \textbf{1.87(0.73)} & * & 2.48 \\
W1-4F1-4 & \textbf{0.76(0.11)} & \textbf{0.88(0.08)} & * & 1.24 & \textbf{0.57(0.14)} & \textbf{0.87(0.07)} & * & 2.74 & \textbf{3.09(0.34)} & 2.02(0.99) & * & 1.44 \\
\bottomrule
\end{tabular}
\end{adjustbox}
\label{tab:imu90}
\end{table*}

\begin{table*}[htbp]
\centering
\caption{Statistical Results for Classification accuracies EMG vs (Accelerometer + Gyroscope + Magnetometer) in 180 degree position.}
\begin{adjustbox}{width=\textwidth}
\begin{tabular}{ l c c c c c c c c c c c c }
\toprule
 \textbf{placement} & \multicolumn{4}{c}{\textbf{LDA}} & \multicolumn{4}{c}{\textbf{SVM}} & \multicolumn{4}{c}{\textbf{DBI}} \\
\cmidrule(lr){2-5} \cmidrule(lr){6-9} \cmidrule(lr){10-13} 
&EMG & IMU & $p$ & $d$ & EMG & IMU & $p$ & $d$ & EMG & IMU & $p$ & $d$  \\
\midrule
W1W2 & 0.43(0.04) & 0.73(0.06) & * & 5.64 & 0.13(0.03) & 0.75(0.08) & * & 10.58 & 6.29(1.19) & \textbf{1.99(0.81)} & * & 4.24 \\
W3W4 & 0.38(0.03) & 0.70(0.05) & * & 7.11 & 0.16(0.02) & 0.73(0.05) & * & 14.10 & 6.19(1.47) & 2.18(0.58) & * & 3.58 \\
F1F2 & 0.32(0.03) & 0.69(0.09) & * & 5.54 & 0.12(0.02) & 0.73(0.10) & * & 8.79 & 5.62(1.46) & 2.35(0.81) & * & 2.77 \\
F3F4 & 0.45(0.05) & 0.72(0.07) & * & 4.43 & 0.14(0.04) & 0.76(0.06) & * & 11.48 & 5.89(0.65) & 2.44(0.69) & * & 5.14 \\
W1-4 & 0.64(0.07) & 0.83(0.05) & * & 3.01 & 0.35(0.08) & 0.83(0.04) & * & 8.00 & 4.78(0.84) & 2.04(0.77) & * & 3.41 \\
F1-4 & 0.57(0.07) & 0.83(0.06) & * & 3.93 & 0.27(0.07) & 0.83(0.06) & * & 8.48 & 4.91(1.06) & 2.16(0.65) & * & 3.13 \\
W1-4F1-4 & \textbf{0.78(0.06)} & \textbf{0.89(0.03)} & * & 2.37 & \textbf{0.60(0.06)} & \textbf{0.87(0.04)} & * & 5.77 & \textbf{3.83(0.77)} & 2.09(0.86) & * & 2.13 \\
\bottomrule
\end{tabular}
\end{adjustbox}
\label{tab:imuleg}
\end{table*}

\subsection{Optimal Placement}
Our results indicate that sensor placement is a critical factor for achieving optimal classification performance with IMU sensors. When using only accelerometer data, placing sensors on either the anterior or posterior forearm, or on the wrist, yields comparable accuracies. However, the number of sensors and the extent of coverage across both anterior and posterior sides of the arm play a more significant role. As shown in Figure ~\ref{fig:tables} configurations such as W1–W4 or F1–F4 result in higher accuracies compared to those covering only one side of the wrist (e.g., W1W2) or forearm.

This pattern is also observed when using magnetometer data. Furthermore, when incorporating all nine axes of the IMU (accelerometer, gyroscope, and magnetometer), results show that placing sensors on the anterior wrist yields better performance than the posterior wrist. Similarly, posterior forearm placements tend to outperform anterior forearm placements in certain models. Nevertheless, the general principle remains consistent: increasing the number of sensors to cover a broader range of muscle groups enhances classification accuracy. Specifically, deploying four sensors around the wrist or forearm improves accuracy by approximately $12\%$, while using eight sensors to cover both regions can lead to improvements of up to $20\%$ compared to configurations with only two sensors on one side of the wrist.

\section{Conclusion}
\label{sec:conclusion}
The findings of this study highlight the potential of using IMU signals for static gesture classification, achieving an average accuracy of $89 \pm 3\%$. Our results indicate that while the anterior part of the wrist and the posterior part of the forearm are more effective locations for IMU sensor placement, sensor quantity and coverage across different regions of the arm play an even more critical role in improving performance. Furthermore, although the wrist contains significantly less muscle mass than the forearm, making it challenging for surface EMG to capture high-quality signals, IMU sensors are capable of detecting subtle movements of muscles and tendons in the wrist area, providing valuable information for gesture recognition. 


\section{Future Work}
\label{sec:limitations}

This study demonstrated that IMU can serve as an alternative to EMG for static gesture recognition at the wrist and forearm, opening up several opportunities for future exploration. 
While this work focused on static gestures in controlled settings, extending it to dynamic gestures performed in more naturalistic environments is essential for practical applications. Evaluating robustness under conditions such as arm movement, sensor displacement, and long-term use will provide insights into the feasibility of IMU-based systems outside the laboratory. To further bridge the gap to daily real-world use, a promising direction is the development of an IMU-based wristband with embedded real-time gesture detection, designed for simplicity and reliability. Such a device could seamlessly connect to prosthetic hands or other electronics and transmit control signals via BLE.
Together, these directions represent important steps toward making IMU-based gesture recognition a practical and accessible solution for real-world human–machine interaction.

\bibliographystyle{IEEEtran}
\setlength{\itemindent}{0pt} 
\bibliography{references}
\end{document}